\documentclass[aps,
 prl,
 reprint,
superscriptaddress,
nofootinbib,
nobibnotes,
amsmath,amssymb,
nolongbibliography,
noeprint,
floatfix
]{revtex4-2}
\usepackage{caption}
\usepackage{soul}
\usepackage{ragged2e} 
\usepackage{float}
\usepackage{graphicx}
\usepackage{mathrsfs}  
\usepackage{amsthm}
\usepackage{physics}
\usepackage[pdftex,colorlinks=true,linkcolor=blue,citecolor=blue,urlcolor=blue]{hyperref}

\usepackage{tikz}
\makeatletter
\renewcommand{\@makecaption}[2]{%
  \vskip\abovecaptionskip
  \begin{minipage}{\linewidth} 
  \small
  \parbox{\linewidth}{\justify #1. #2} 
  \end{minipage}
  \vskip\belowcaptionskip
}
\makeatother
\usepackage{placeins}
\usepackage{wrapfig}
\usepackage{hyperref}

\newcommand{\sect}[1]{\par\textit{#1}---}

\newcommand{\E}{\mathscr{E}}
\newcommand{\AM}{\mathscr{L}}

\newcommand{\K}{\mathscr{K}}

\newcommand{\consts}{\mathcal{C}}

\begin{document}

\preprint{APS/123-QED}
\title{Pitching Cosmic Curveballs: Environmental Effects on Extreme-Mass-Ratio \\ Inspirals with Spinning Secondaries}

\author{Leif Lui}
\affiliation{Beijing Institute of Mathematical Sciences and Applications, Beijing 101408, China}

\author{Lisa V. Drummond}
\affiliation{Department of Physics and TAPIR, California Institute of Technology, Pasadena, CA}

\author{Alejandro Torres-Orjuela}
\email[Corresponding author: ]{atorreso@bimsa.cn}
\affiliation{Beijing Institute of Mathematical Sciences and Applications, Beijing 101408, China}

\date{\today}

\begin{abstract}
Much like the aerodynamic deflection of a spinning curveball, a rotating secondary in an extreme-mass-ratio inspiral (EMRI) experiences Magnus and lift forces, in addition to the standard drag force, when traversing a gaseous environment. We present the first framework that incorporates these specific spin-coupled environmental effects (EEs) into the evolution of EMRI. Over the multi-year observation windows of space-based gravitational-wave (GW) detectors, these interactions imprint a unique, distinguishable dephasing signature on the signal. Crucially, a Fisher matrix analysis reveals that gas drag breaks the fundamental vacuum-projection degeneracy between the secondary's spin magnitude and inclination, thereby tightening parameter constraints. Thus, accounting for EEs is not merely a modeling necessity, but could potentially be a powerful tool for enhancing the detectability of the secondary's intrinsic spin, and could serve as a novel probe of accretion flows harboring massive black holes.
\end{abstract}

\maketitle
\begin{figure}[t]
    \centering
    \includegraphics[width=\linewidth]{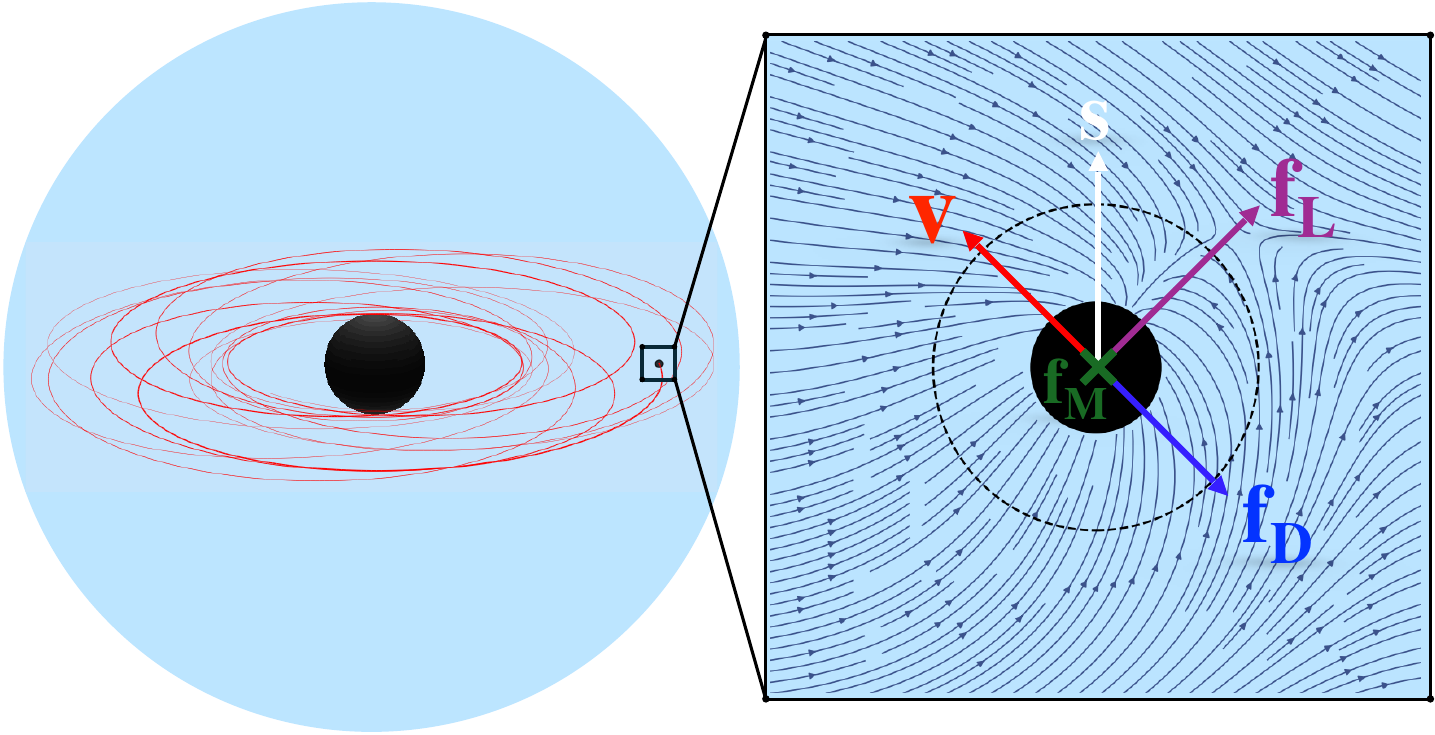}
    \caption{Aerodynamic forces imparted onto a spinning secondary of an EMRI submerged in gas. The drag force (blue arrow) is anti-parallel to the secondary's velocity vector (red arrow). The Magnus force (cross) is (anti-)aligned with the vector normal to the plane spanned by the velocity vector and the spin vector (white arrow). The lift force (purple arrow) is orthogonal to both the drag and Magnus forces. The black vector field lines represent the stream lines of the gas.}
    \label{EMRI_Aerodynamics_Diagram}
\end{figure}

\sect{Introduction}Just as a spinning curveball drags the surrounding air to create a lateral Magnus force that sharply deflects its path, spinning black holes (BHs) submerged in astrophysical environments experience analogous aerodynamic forces. As illustrated in FIG.~\ref{EMRI_Aerodynamics_Diagram}, a moving BH creates a trailing overdensity in the surrounding gas, causing a gravitational tug, or drag force $f_\mathrm{D}$~\cite{Chandrasekhar_1943, Ostriker_1999, Escala_2003, Kim_2007, Baurausse_2008,  Kocsis_2011, Yunes_2011, Derdzinski_2021, Traykova_2021, Traykova_2023, Zwick_2025}. A spinning BH also experiences analogous Magnus, $f_{\mathrm{M}}$, and lift forces, $f_\mathrm{L}$~\cite{Costa_2018, Wang_2024, Dyson_2024, Karydas_2026}. The Magnus force arises from a spin-induced overdensity in the top half of a hemisphere whose equator lies in the plane spanned by the velocity and spin vectors. The lift force is a result of a density asymmetry caused by the breaking of axisymmetry in the gravitational wake.

These environmental effects (EEs) can influence the orbit of binary BH mergers submerged in gaseous media, and the development of future space-based gravitational wave (GW) detectors, like LISA~\cite{Amaro_Seoane_2017, babak_gair_2017, Barack2007, Amaro_Seoane_2018, LISA_2022b, LISA_2022c, LISA_2023, LISA_2024}, TianQin~\cite{TianQin_2015, TianQin_2021, TianQin_2024, Torres-Orjuela_2024}, and Taiji~\cite{taiji_2015, Ruan_2018}, provides an exciting opportunity to probe these effects and elucidate the nature of BH environments. One particular source of interest for space-based GW detectors is extreme-mass-ratio inspirals (EMRIs), where a small secondary orbits a massive BH (MBH), forming binaries with small mass ratios of $q=\mu/M\gtrsim 10^{-4}-10^{-7}$~\cite{Barack2007}. The small mass ratio of EMRIs allows the trajectory and GW waveforms to be computed perturbatively using the Teukolsky formalism and gravitational self-force (GSF) methods~\cite{Teukolsky1972, Teukolsky1973, Teukolsky1974, Chandrasekhar_1975, Poisson_1993, Cutler_1993, Apostolatos_1993, Poisson_1993b, Poisson_1995b, MST_1996, MST_1997, Campanelli_1997, Hughes_2000, Lo_2024}.

When modeling EMRIs with spinning secondaries, the extended nature of the body requires including the leading secondary-spin correction. Recent studies suggest that the spin-curvature coupling (SCC)~\cite{Drummond_2023, Drummond_2024, Skoupy_2025, Piovano_2026} and linear-in-secondary-spin (LISS) fluxes~\cite{Skoupy_2023, Piovano_2024, Drummond_2026, Skoupy_2026} could be non-negligible and significantly dephase the GW signals over the long observation time of space-based GW detectors~\cite{Gair_2012, Piovano_2021, Cui_2025, Burke_2024, Drummond_2024}. More recently, there is evidence that the secondary spin could be measurable for eccentric orbits~\cite{Skoupy_2025b}. However, this has not been demonstrated through rigorous data analysis. While EEs, such as drag, mass transfer, and shocks, have been well-studied for EMRIs with non-spinning secondaries~\cite{Baurausse_2008, Dai_2022, Dai_2024, Mitra_2025, Vicente_2025, Lui_2025, Hegade_2025, Hegade_2025b}, literature on spin-coupled EEs is scarce~\cite{Dyson_2024, Karydas_2026}. In this Letter, we provide the first model that incorporates these spin-environment coupling effects into EMRI models with spinning secondaries submerged in gaseous environments, and demonstrate that the secondary spin can be constrained when incorporating these EEs for low-eccentricity EMRIs on equatorial orbits.

\sect{Gravitational Self-Forces and Environmental Effects}Under the pole-dipole approximation, the trajectory of a spinning body in curved spacetime is governed by the Mathisson-Papapetrou-Dixon (MPD) equations~\cite{Papapetrou_1951, Dixon_1970, Mathisson_2010}, which are closed by applying the Tulczyjew-Dixon spin supplementary condition, $S^{\mu\nu}p_{\mu}=0$. While recent progress has yielded analytic solutions to the MPD equations in Kerr spacetime~\cite{Witzany_2019, Witzany_2024, Witzany_2025, Ramond_2025}, it was notably demonstrated in Ref.~\cite{Skoupy_2025} that these equations are separable in the LISS approximation. We leverage these solutions to compute the trajectory of a massive spinning particle and subsequently evolve the inspiral using the fluxes associated with GW radiation and EEs.

For a primary BH of mass $M$ and spin $a$ described by the Kerr metric~\cite{Kerr_1963}, the orbital dynamics of a spinning secondary are completely characterized, to LISS~\cite{Pound2008, Gair2011, Skoupy_2021, Skoupy_2025b, Drummond_2026}, by the spin-corrected constants of motion $\consts\in\{\E, \AM, \K\}$~\cite{Skoupy_2025, Skoupy_2025b, Skoupy_2026}. When considering external forces, $f^{\mu}$, in Boyer-Lindquist coordinates, $(t,r,\vartheta,\varphi)$, exerted on a spinning particle along its worldline, these constants of motion evolve as
\begin{equation}\label{exact_flux}
    \frac{\dd \E}{\dd\lambda}=-\hat{g}_{t\mu}f^{\mu},\quad \frac{\dd \AM}{\dd\lambda}=\hat{g}_{\varphi\mu}f^{\mu},\quad \frac{\dd \K}{\dd\lambda}=2\K_{\mu\nu}f^{\mu}u^{\nu},
\end{equation}
where $\hat{g}_{\mu\nu}$ is the background metric, $\K_{\mu\nu}$ is the Killing tensor~\cite{Carter_1968}, $u^{\mu}$ is the 4-velocity of the secondary, and $\lambda$ is the Mino-Carter time~\cite{Mino2003}.
\begin{figure*}[ht!]
    \centering
    \includegraphics[width=\linewidth]{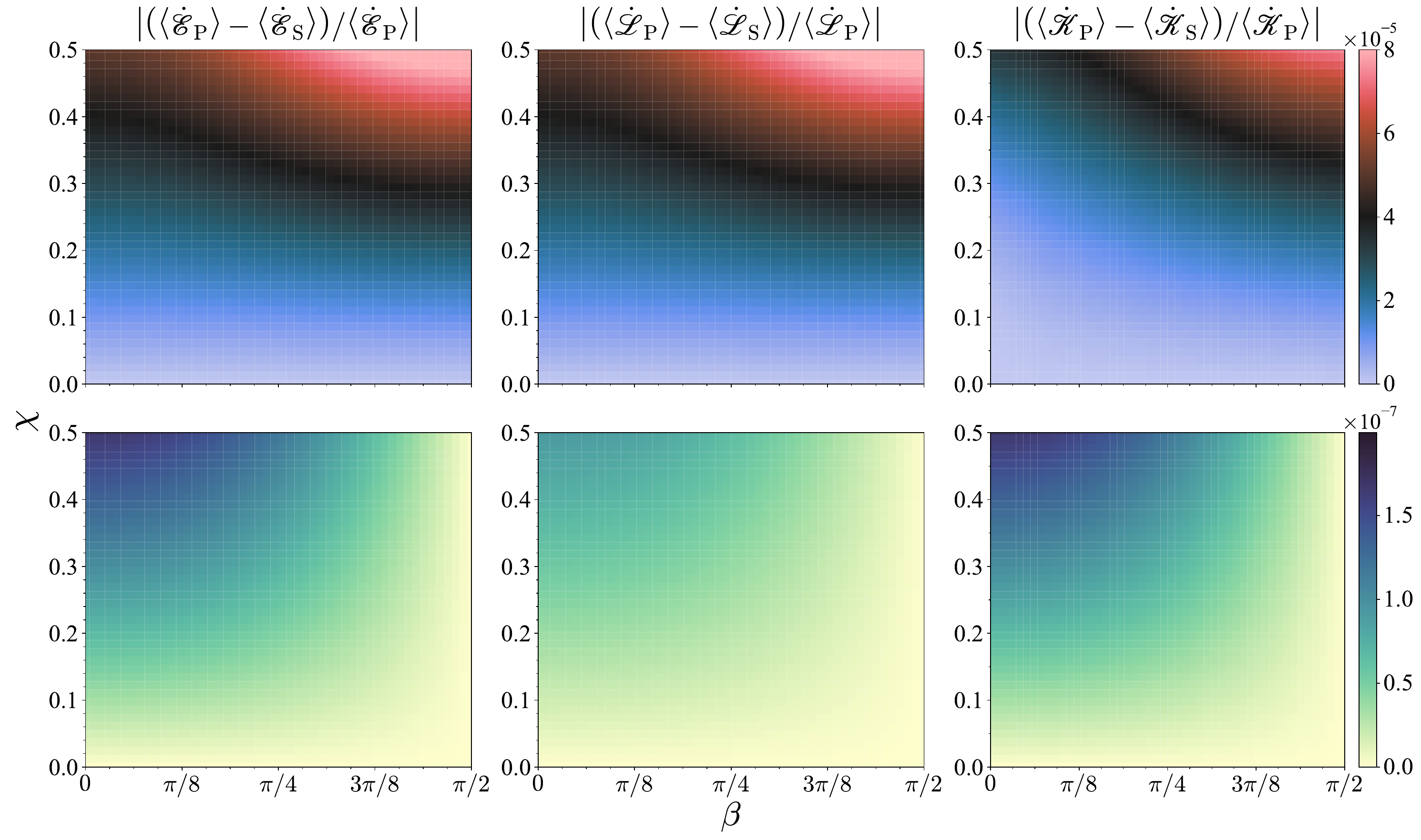}
    \caption{Relative difference in aerodynamic (top) and GW (bottom) fluxes of $(\E, \AM, \K)$ between a non-spinning point mass ($\mathrm{P}$) and a spinning dipole ($\mathrm{S}$). Axes denote the secondary's spin magnitude $\chi$ and the spin-velocity angle $\beta$. The parameters $(a, p, e, x_I) = (0.55M, 12.63M, 0.5, 0.5)$ and $q=10^{-5}$ are fixed.}    \label{Aerodynamic_Flux_comparison}
\end{figure*}
In a vacuum, spinning EMRIs evolve through GSFs and SCC. However, using the full osculating GSFs directly is impractical for studying long-term orbital evolution. At each step, the full osculating GSFs depend on the orbital parameters $(a,p,e,x_I)$ and phases $(q_t,q_r,q_\vartheta,q_\varphi)$, requiring a summation over a large spectrum of Fourier modes $(\ell,m,n,k)$. For practical computation over long timescales, the convention is to reduce the parameter space by computing the flux averaged over the orbital phases~\cite{Mino2003, Drasco_2004, Drasco_2006, Isoyama2019, Sago2026, Yin_2026}
\begin{equation}\label{averaged_fluxes}
    \langle\dot{\consts}\rangle=\frac{q}{(2\pi)^2}\int^{\pi}_{-\pi}\int^{\pi}_{-\pi}\frac{\dot{\tau}}{r^2+a^2\cos^2\vartheta}\frac{\dd \consts}{\dd\lambda}\dd q_r\dd q_{\vartheta},
\end{equation}
where $\tau$ is the proper time and the overdot denotes the derivative with respect to the coordinate time, $t$.

\begin{figure*}
    \centering
    \includegraphics[width=\linewidth]{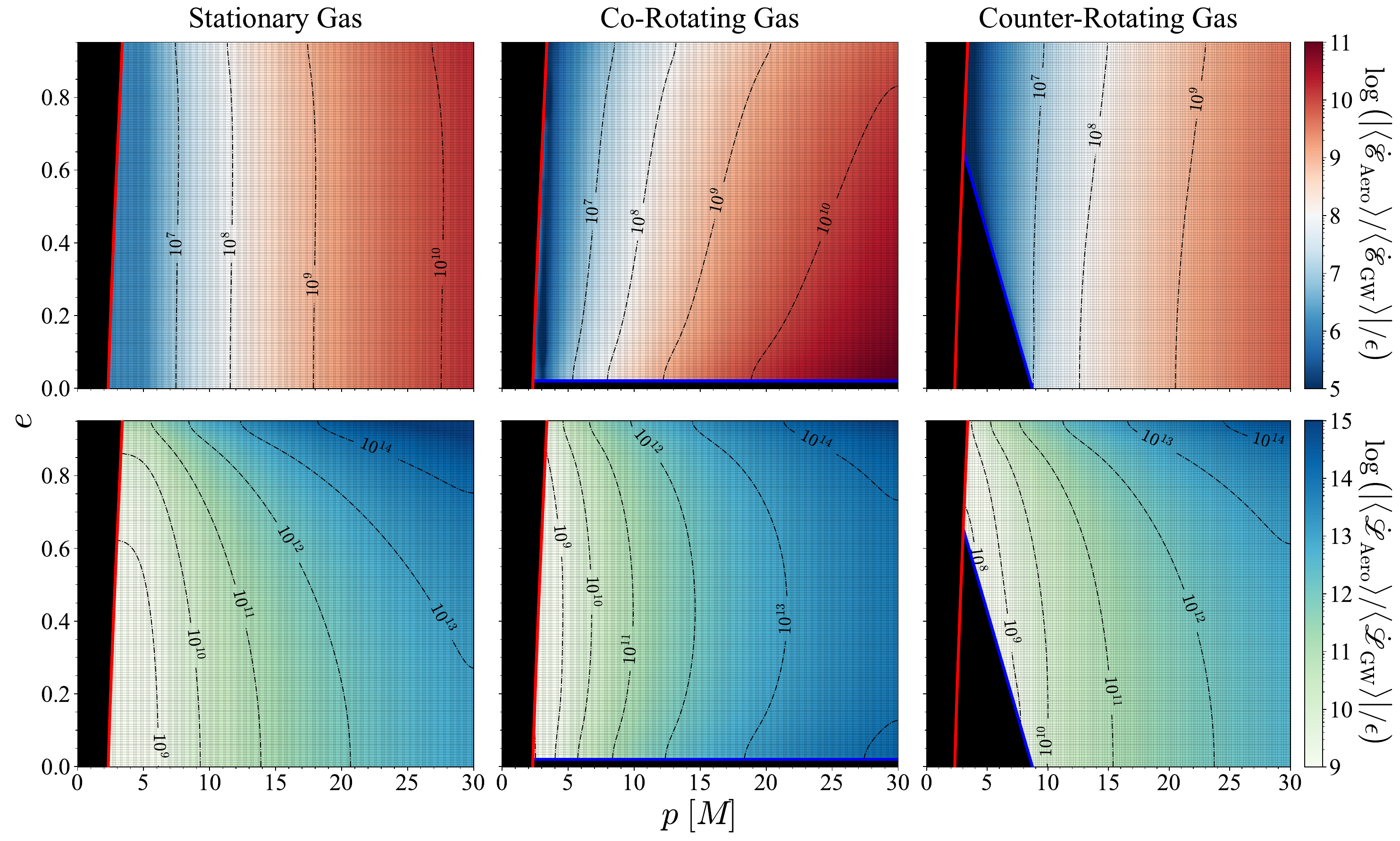}
    \caption{Contour plots of the ratio of aerodynamic drag to GW fluxes $\dot{\E}$ (top row) and $\dot{\AM}$ (bottom row), scaled by the density perturbation parameter $\varepsilon=\rho_{\mathrm{g}}/\rho_{\bullet}$. This is evaluated for the cases where the gas is stationary (first column), co-rotating (second column), and counter-rotating (third column), with respect to the orbital direction. The red line denotes the location of the last stable orbit of the prograde EMRI, and the blue line shows regions where the fluxes due to EEs are zero or divergent. In all panels, the EMRI orbits are confined to the equatorial plane with a non-spinning secondary and primary spin of $a=0.9M$. }
    \label{non-spinning_flux_comparison}
\end{figure*}

In this Letter, we confine our study to adiabatic (0PA) GSF to first order in $q$ (1SF). This is because, for the binary parameters and environmental densities considered, the EEs appear to dominate over pure spinning secondary effects and, in some cases, even the 0PA-1SF term (cf. FIG.~\ref{Aerodynamic_Flux_comparison} and \ref{non-spinning_flux_comparison}). Thus, we expect them to also dominate over the first post-adiabatic (1PA) conservative corrections~\cite{van_de_Meent_2018} just like the 0PA-1SF term. Given the dominance of EEs over $\mathcal{O}(q^2)$ effects for the mass ratios considered here, the adiabatic 0PA-approximation is sufficient for the present analysis, bypassing the need for currently incomplete 1PA-2SF Kerr corrections~\cite{Pound_2017, Warburton_2021, Wardell_2023, Burke_2024, Spiers_2024, Wei_2025, Matthews_2026, Lewis_2026, Upton_2026}. We interpolate the average fluxes using the \texttt{FastEMRIWaveforms} (\texttt{FEW}) dataset~\cite{Chua_2019, Chua_2021, Chua_2021b, Chapman-Bird_2025}, extending it to higher eccentricities ($e \leq 0.95$) with the \texttt{pybhpt} package~\cite{Nasipak_2022, Nasipak_2024}. To guarantee convergence at these high eccentricities, we adopt the dynamic mode-truncation scheme in Ref.~\cite{Drasco_2006} (See End Matters for details). Similarly, we use the \texttt{KerrSpinningFluxes} package~\cite{Skoupy_2023, Skoupy_2025b} to include 1PA SCC and LISS GW flux corrections, subjecting them to the same truncation scheme. 

To evolve the EOM of EMRIs with a spinning secondary moving through a gaseous medium, we first construct a set of tetrads $e_{A}^{\mu}$ that are parallel transported along a forced timelike Kerr geodesic. In the tetrad frame, we calculate the aerodynamic forces in a particle-like medium using the results in Ref.~\cite{Dyson_2024} (see End Matters for more details). To obtain the force in the Boyer-Linquist frame, we contract the forces in the tetrad frame with the $e_{A}^{\mu}$. That is, $f^{\mu}_{\mathrm{EE}}=e^{\mu}_Af^A_{\mathrm{EE}}$, where $f^{\mu}_{\mathrm{EE}}$ are the aerodynamic forces in the BL frame, and $f^A_{\mathrm{EE}}$ are the aerodynamic forces in the tetrad frame. In Cartesian coordinates, $f^A_{\mathrm{EE}}=(f_\mathrm{M}, f_\mathrm{L}\cos\beta-f_\mathrm{D}\sin\beta,f_\mathrm{D}\cos\beta+f_\mathrm{L}\sin\beta)$, where $f_\mathrm{D}$, $f_\mathrm{L}$, $f_\mathrm{M}$, are the drag, lift and Magnus forces as depicted in FIG.~\ref{EMRI_Aerodynamics_Diagram}, respectively, and $\beta$ is the angle between the secondary's velocity and spin vectors. Finally, we can compute the orbit-averaged fluxes associated with LISS EEs by using $f^{\mu}_{\mathrm{EE}}$ in Eq.~\eqref{exact_flux} and integrating the flux over the orbital phases using Eq.~\eqref{averaged_fluxes}.

\sect{Spin-Coupled Environmental Effects on EMRIs}To understand the spin-coupling terms, we perform a 3-parameter expansion on the force $f^{\mu}$ along the geodesic using the mass ratio $q$, $\varepsilon=\rho_{\mathrm{g}}/\rho_{\bullet}$, and $s=\chi q$ as expansion parameter, where $\varepsilon$ compares the gas density $\rho_{\mathrm{g}}$ with the BH density $\rho_{\bullet}$~\cite{Dyson_2025}, and $s$ scales the secondary spin $\chi$ with the mass ratio $q$. Keeping terms linear in $q$, $s$, and $\varepsilon$
\begin{equation}\label{geodesic_forces} f^{\mu}=q\left[f^\mu_{\mathrm{GSF}}+sf^\mu_{\mathrm{LISS}}+\varepsilon\left(f^\mu_{\mathrm{\text{NS-EE}}}+sf^\mu_{\mathrm{S\text{-}EE}}\right)\right].
\end{equation}
The terms $f^\mu_{\mathrm{GSF}}$ and $f^\mu_{\mathrm{LISS}}$ correspond to vacuum corrections (0PA-1SF and LISS corrections, respectively), while $f^\mu_{\mathrm{S/NS\text{-}EE}}$ are the aerodynamic forces from the environment due to a spinning/non-spinning secondary, respectively. Therefore, $f^\mu_{\mathrm{LISS}}$ and $f^\mu_{\mathrm{S\text{-}EE}}$ represent the spin-coupling corrections.

Recent studies show that LISS terms, $f^\mu_{\mathrm{LISS}}$, which encompass spin-curvature coupling and first-order spin self-forces, are subdominant to the leading 0PA-1SF~\cite{Skoupy_2021, Skoupy_2023, Drummond_2024}. While a misaligned secondary spin formally induces precession of both the spin vector and orbital plane, these dynamics only enter at higher PA orders~\cite{Skoupy_2021, Piovano_2021, Drummond_2024}. Thus, we compute leading-order secular fluxes using a fixed equatorial orbit with a constant spin projection.

To quantify the secondary's extended nature, FIG.~\ref{Aerodynamic_Flux_comparison} plots the relative difference, $|(\langle \dot{\mathcal{C}}_\mathrm{P}\rangle-\langle \dot{\mathcal{C}}_\mathrm{S}\rangle ) / \langle \dot{\mathcal{C}}_\mathrm{P}\rangle|$, between fluxes from a non-spinning point mass ($\mathrm{P}$) and a spinning dipole ($\mathrm{S}$). The top panel compares the purely mass-driven EEs ($f^\mu_{\mathrm{NS\text{-}EE}}$) against the spin-environment couplings ($f^\mu_{\mathrm{S\text{-}EE}}$). The bottom panel compares the 1PA GW spin corrections ($f^\mu_{\mathrm{LISS}}$)~\cite{Skoupy_2021, Skoupy_2023, Drummond_2024, Skoupy_2026} against the leading 0PA GW fluxes ($f^\mu_{\mathrm{GSF}}$)~\cite{Mino2003, Drasco_2004, Drasco_2006, Isoyama2019}.

For the case with EEs, we see that the deviation between the spinning and non-spinning fluxes grows at higher spin magnitudes $a_s$ and misalignment angles $\beta$. This is expected, as a higher $a_s$ amplifies the Magnus and lift forces, and larger $\beta$ values correspond to a more significant breaking of axisymmetric fluid flow. The converse effect is observed in the pure GW case, where the flux deviation decreases as $\beta$ increases. This is because the velocity direction is orthogonal to the orbital plane, meaning that $\beta=0$ corresponds to $a_s$ being perpendicular to the orbital angular momentum. More interestingly, we find that $sf^\mu_{\mathrm{S\text{-}EE}}/f^\mu_{\mathrm{NS\text{-}EE}}\sim 10^{-5}$ and $sf^\mu_{\mathrm{LISS}}/f^\mu_{\mathrm{GSF}}\sim 10^{-7}$. This suggests that when EEs are comparable to the GSF, their interplay may exacerbate orbital deviations in EMRIs with spinning secondaries. Consequently, accounting for EEs in GW signal modeling could provide improved constraints on the secondary's spin. As a result, it is prudent to compare the 0PA GW fluxes with the leading-order EEs (pure drag) to identify the regime in which they are of similar magnitude.

In FIG.~\ref{non-spinning_flux_comparison}, we plot the ratio between aerodynamic and GW fluxes for $\chi=0$ across the $(p, e)$-plane for EMRIs in the mHz GW band. The ratio $|\dot{\consts}_{\mathrm{EE}}/\dot{\consts}_{\mathrm{GW}}|$ is scaled by the perturbation parameter $\varepsilon$, representing the environment density surrounding the MBH. For water, $\varepsilon\sim 10^{-5}$, which makes the fluxes dominant due to EEs for most regions of the parameter space. However, the gas density near the last stable orbit (LSO) of a MBH in accretion disks is highly variable, ranging from $10^{-12}$--$10^{-1}\;\mathrm{g\cdot cm}^{-3}$~\cite{Mallick_2018, Tomsick_2018, Jiang_2019, Fabian_2020, Xu_2021, Jiang_2022, Mallick_2022, Ding_2024, Madathil-Pottayil_2024}, which corresponds to $\varepsilon\sim10^{-17}$--$10^{-6}$. Even at these densities, the scaled ratios of $|\dot{\consts}_{\mathrm{EE}}/\dot{\consts}_{\mathrm{GW}}|\sim10^{7}$--$10^{11}$ suggest that EEs can, in some cases, dominate over the 0PA-1SF across the relevant regions of the parameter space. This dominance of gas drag at larger radii and low relative velocities is consistent with previous studies on EMRI with EEs~\cite{Baurausse_2008, Kocsis_2011, Yunes_2011, Derdzinski_2021}. However, while those prior works focused purely on the isotropic $f_{\mathrm{D}}\sim1/v^2$ deceleration, our framework demonstrates that incorporating the secondary's spin breaks this symmetry, introducing lateral aerodynamic forces that distinctly alter the secular evolution. We focus on EMRIs with low eccentricities, as compact objects that form in or are captured by accretion disks~\cite{Peng_2023, Peng_2025} are expected to experience rapid orbital circularization due to these dissipative EEs~\cite{Duque_2026, Zeng_2026, Hegade_2025, Hegade_2025b}. 

The specific impact depends on the orbital direction relative to the gas flow. When the gas is stationary, the aerodynamic drag provides a substantial correction peaking at $\sim 10^{8}$ near the LSO. For co-rotating gas, the steady-state formulation diverges as relative velocity drops to zero, as $f_\mathrm{D}\sim 1/v^2$~\cite{Chandrasekhar_1943}. This divergence is a mathematical artifact, as physical spherical symmetry is restored at absolute rest. We avoid this breakdown and remain within the valid aerodynamic regime by strictly restricting our analysis to orbits with $e \geq 0.005$. Conversely, for counter-rotating gas, the orbiter follows a prograde trajectory while the gas flows in a retrograde direction. Both co- and counter-rotating cases 
strongly favor circular orbits, as the relative velocity between the gas and the orbiter is lower for low $e$. However, in the counter-rotating case, because the accretion disk is truncated when $r\leq r_{\mathrm{LSO}}(-a,e,x_I)$, the prograde orbiter emits purely GWs in these voided regions as it continues its inspiral. Another interesting feature is that $|\dot{\consts}_{\mathrm{EE}}/\dot{\consts}_{\mathrm{GW}}|$ for the counter-rotating case is generally smaller than the stationary and co-rotating cases. This is because the secondary's velocity is nearly anti-parallel to the gas velocity. This increases the relative velocity, reducing the fluxes due to drag. In all three cases, in the strong-gravity region, $|\dot{\consts}_{\mathrm{EE}}/\dot{\consts}_{\mathrm{GW}}| \sim 10^7$, implying that EEs can be of similar order to the 0PA GW fluxes as shown in dense accreting environments. 

\begin{figure}[t]
    \centering
    \includegraphics[width=\linewidth]{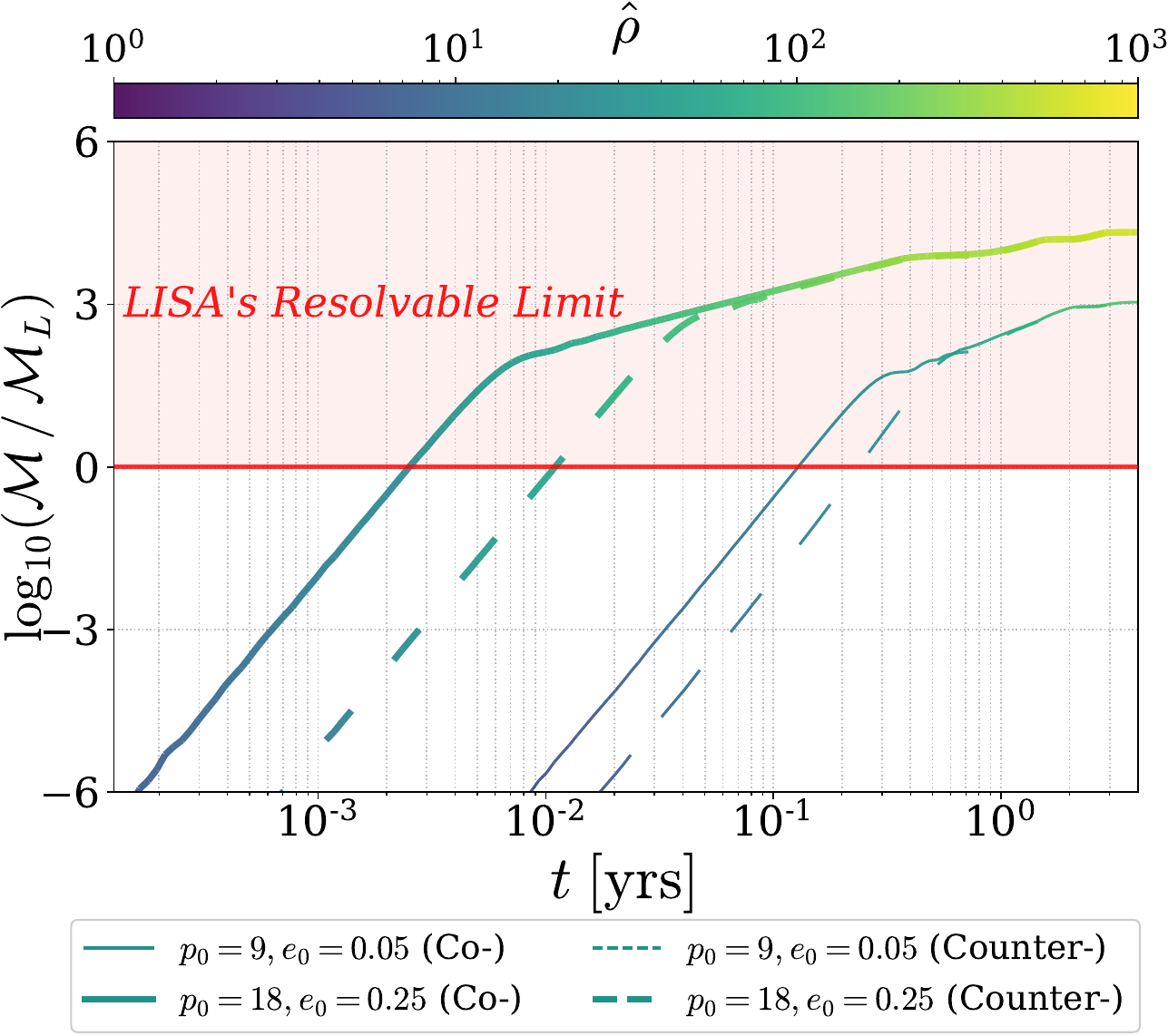}
    \caption{Evolution of the mismatch, $\mathcal{M}$, between an EMRI with a spinning secondary with and without EEs, evaluated over LISA's observation time ($4\,{\rm yr}$). The color axis shows the evolution of the signal-to-noise ratio (SNR), $\hat{\rho}$, at a luminosity distance of $d_\mathrm{L}=500\;\mathrm{Mpc}$. The mismatch is normalized by the mismatch that satisfies the Lindblom criterion~\cite{Lindblom_2008}, $\mathcal{M}_L=D/(2\hat{\rho}^2)$, where $D=15$ is the number of source parameters. The thickness of lines represents EMRIs with different initial orbital parameters, namely $(p_0,e_0)=(9M,0.05)$ and $(p_0,e_0)=(18 M,0.25)$ for the thin and thick lines, respectively. The solid/dashed line styles denote whether the EMRI is co-/counter-rotating with respect to the gas. The parameters $a=0.9$, $M=2.5\times10^6M_{\odot}$, $\chi=0.25$, $q=1.2\times10^{-5}$, $\beta=\pi/4$, $x_{I0}=1$, and $\varepsilon=10^{-11}$ ($\rho_{\mathrm{g}}\approx6.25\times 10^{-6}\;\mathrm{g\cdot cm^{-3}}$) are fixed, and the SNR and mismatches are calculated using LISA's noise curve~\cite{Cornish_2003, Babak_2007, Klein_2016, Amaro_Seoane_2023, LISA_2024}.}
    \label{mismatch_evolution}
\end{figure}
In FIG.~\ref{mismatch_evolution}, we compute the evolution of the mismatch between EMRI waveforms with and without EEs for the late ($p_0=9M,\;e_0=0.05$) and early ($p_0=18M,\;e_0=0.25$) inspiral cases. To determine whether the GW signal of an EMRI with EEs can be differentiated from EMRIs in vacuum, we normalize the mismatch with $\mathcal{M}_L=D/(2\hat{\rho}^2)$. $\mathcal{M}_L$ satisfies the Lindblom criterion~\cite{Lindblom_2008}, which is a necessary, but not sufficient condition for distinguishability. We find that the mismatch grows more rapidly for cases far from the LSO, and satisfies the Lindblom criterion earlier in the inspiral. This implies that EEs become more prevalent, as they dominate over the GSFs in early inspirals. Moreover, we compare the cases when the gas is co-/counter-rotating with respect to the secondary. We find that in general, the mismatch evolves more rapidly when the secondary is co-rotating with the gas. For nearly circular, co-rotating orbits, the relative velocity between the secondary and the gas is minimized, providing the secondary more time to accumulate a dense gravitational wake, resulting in enhanced EEs. This again agrees with FIG.~\ref{Aerodynamic_Flux_comparison} and~\ref{non-spinning_flux_comparison}, and suggests space-based detectors have a better chance of probing EEs in early EMRIs where the matter in astrophysical environments (i.e., accretion disks, dark matter halos, etc.~\cite{Dyson_2026, Karydas_2026, Mach_2026, Hegade_2025, Hegade_2025b}) is co-rotating with the secondary.

\sect{Constraints on the Secondary's Spin}The significant dephasing and cumulative mismatches due to EEs observed in FIG.~\ref{mismatch_evolution} suggest that EEs do not merely mask the underlying vacuum signal, but rather enhance the information content available for parameter estimation (PE). Specifically, the high signal-to-noise ratios (SNRs) achievable for EMRIs over multi-year LISA observations facilitate a precise mapping of the orbital trajectory, potentially enabling more stringent constraints on the secondary's spin parameters $\chi$ and $\beta$.

To quantify the precision of these constraints, we employ a Fisher Matrix (FM) analysis. The FM, $\Gamma_{ij} = ( \partial_i h | \partial_j h )$, provides a local measure of the curvature of the likelihood surface around the true parameter values, where the marginalized $1\sigma$ uncertainties are given by the square root of the diagonal elements of the covariance matrix $\Sigma = \Gamma^{-1}$~\cite{Finn_1992, Cutler_1994, Vallisneri_2008}. In the high-SNR regime, the likelihood surface is well-approximated by a multivariate Gaussian. While full Bayesian parameter estimation (PE) would provide a more complete mapping of the posterior distribution~\cite{Veitch_2010, Veitch_2015, Meidam_2018, Feng_2019, Pizzati_2022}, such an approach is currently infeasible due to the high computational cost of generating forced inspirals and Teukolsky-based waveforms, as the phase rapidly oscillates over the slow orbital timescales~\cite{Lynch_2021} and Bayesian PE requires millions of sequential evaluations to integrate the flux-driven evolution over the lengthy EMRI lifespan~\cite{Chua_2021, Katz_2021, Chapman-Bird_2025}. When computing the FM, we first rescale the matrix via Jacobi preconditioning, and perform the inversion using 100-digit precision in \texttt{mpmath}~\cite{Vallisneri_2008}. This is necessary because the EMRI FM is typically ill-conditioned, as the wide dynamic range of parameters produces derivatives that spans over many orders of magnitude, so double-precision inversion is dominated by round-off error~\cite{Piovano_2021}. Finally, to restrict our analysis to physically valid configurations, we inject independent Gaussian priors into the information matrix corresponding to the theoretical maximum bounds of the secondary spin magnitude ($\sigma_{\chi} = 1$), and its inclination angle ($\sigma_{\beta} = \pi/2$)~\cite{Poisson_1995, Piovano_2021}. We note that $\sigma_{\chi} = 1$ reflects the Kerr BH bound and does not apply to a neutron star (NS) or white dwarf (WD) secondary. NS spin, however, is limited by the mass-shedding limit to values near the Kerr BH bound~\cite{Piovano_2020}, whereas for WDs, their large radius makes tidal heating and mass transfer dynamically more important during inspirals~\cite{Datta_2024, Yang_2026}.
\begin{figure}
    \centering
    \includegraphics[width=\linewidth]{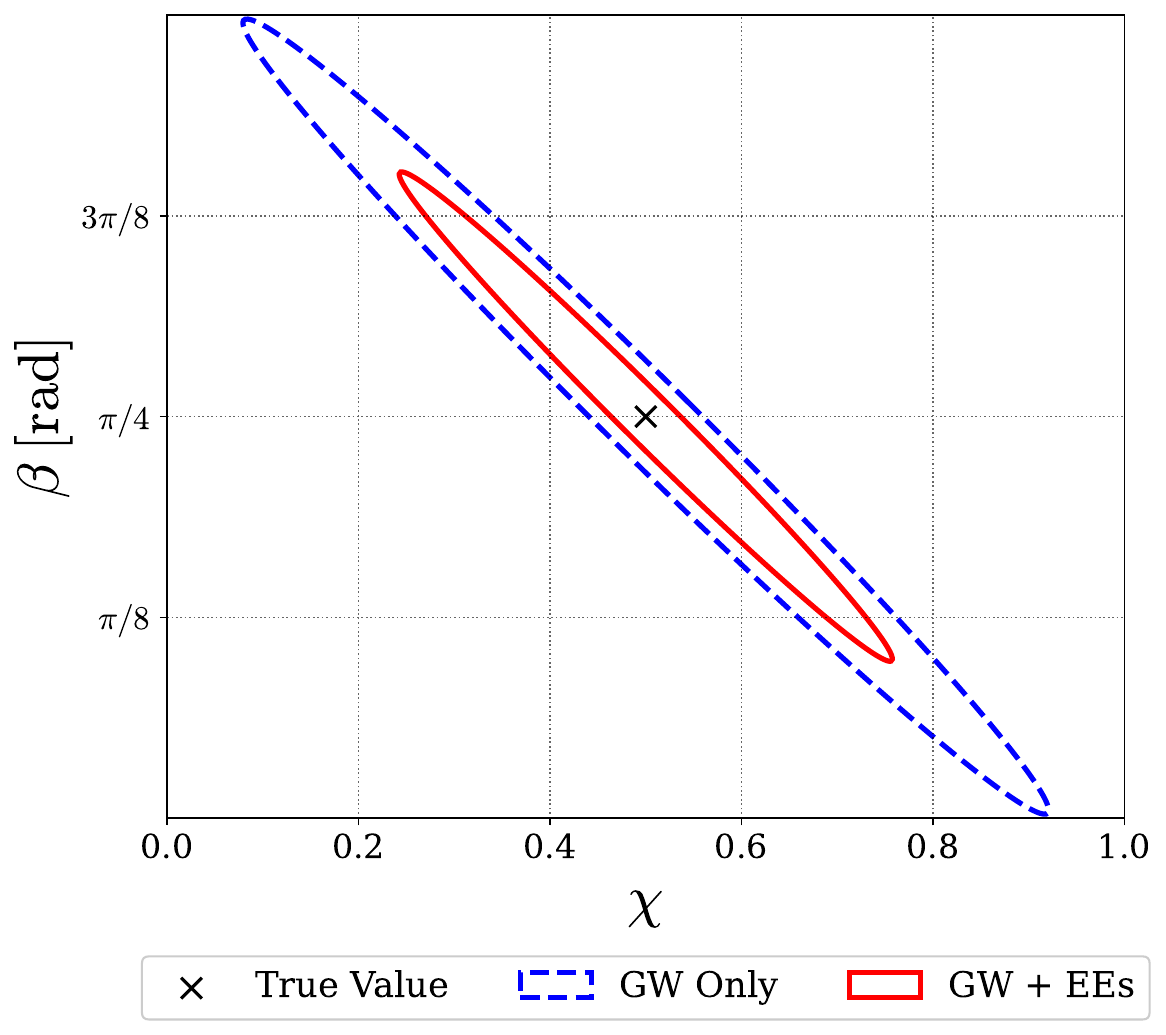}
    \caption{Projected $1\sigma$ Fisher constraints on the secondary's spin dimensionless magnitude $\chi$ and inclination angle $\beta$ for a 4-year EMRI observation ($M=10^6M_{\odot}$, $q = 3 \times 10^{-5}$, $a = 0.9$, $p_0 = 15M$, $e_0 = 0.25$). Contours compare pure vacuum evolution (blue dashed) with a co-rotating accretion disk environment (red solid) with gas density $\varepsilon=7.5\times10^{-13}$ ($\rho_{\mathrm{g}}\approx4.69\times 10^{-7}\;\mathrm{g\cdot cm^{-3}}$), around the true injected values (black cross), $\chi=0.5$ and $\beta=\pi/3$. }
    \label{Fisher_plot}
\end{figure}

Our FM analysis in FIG.~\ref{Fisher_plot} demonstrates that EEs can improve the secondary spin measurement. In vacuum, orbit-averaged GW fluxes (which exclude the secondary's spin precession) depend strictly on the parallel spin projection, $s_{\parallel} \propto \chi \sin\beta$. Consequently, $\chi$ and $\beta$ remain highly degenerate; their apparent individual uncertainties ($\Delta \chi \approx \pm 0.421$) simply saturate the chosen physical priors. EEs break this degeneracy because aerodynamic gas drag depends intimately on the secondary's relative velocity and spin orientation within the local fluid frame, introducing a distinct functional dependence.

\sect{Discussion and Conclusions}In this Letter, we develop the first framework to incorporate spin-coupled EEs (drag, Magnus, and lift forces) into EMRIs. We demonstrate that in dense astrophysical environments, such as accretion disks, these interactions not only dominate over vacuum SCC but can rival 0PA-1SF corrections. Over the multi-year observation windows of space-based detectors, these effects imprint a distinguishable dephasing signature on the GW signal.

Crucially, our FM analysis suggests that EEs can help break certain parameter degeneracies. In vacuum, the secondary's dimensionless spin $\chi$ and inclination $\beta$ are degenerate, as the orbit-averaged 1PA-1SF-spin fluxes depend exclusively on the projected spin component parallel to the orbital angular momentum~\cite{Witzany_2019, Witzany_2026}. Inclusion of EEs, however, breaks this symmetry, as the fluxes depend intimately on the relative velocity and geometry of the secondary within the local fluid frame. This mechanism disentangles the spin magnitude from its orientation, suggesting that EEs could eventually serve as a useful physical signature for constraining the intrinsic properties of the compact secondary.

We conclude by noting some limitations of our model and discussing potential future work. Our model assumes constant gas density, which serves as a local approximation for stable disks. Given that standard dynamical friction typically dominates over spin-induced EEs, managing uncertainties in these EEs is paramount for robust inference, and should encompass the secondary's dynamical back-reaction on the surrounding fluid~\cite{Dyson_2025, Dyson_2026}. To perform accurate and computationally efficient Bayesian PE for these waveforms, implementing near-identity transformations will also be crucial to consistently incorporate 1PA conservative and oscillatory corrections to the GSFs and EEs, as this could provide higher-order phase information to further refine these spin constraints and resolve any residual projection degeneracy~\cite{van_de_Meent_2018, Van_de_Meent_2018b, Lynch_2021, Lynch_2023, Lynch_2024, Drummond_2024}. Despite these modeling complexities, capturing EEs provides a unique avenue to infer the secondary's intrinsic properties~\cite{Karydas_2025}. Extending EEs frameworks to incorporate magnetohydrodynamics will ultimately transform EMRIs into precision probes of both fundamental physics and BH accretion environments.

\sect{Acknowledgements}We thank Xian Chen, Viktor Skoup\'{y}, Samson Leong, Yang Yang, and Ver\'{o}nica V\'{a}zquez-Aceves for valuable discussions. LL and ATO were supported by the Beijing Natural Science Foundation (No. IS25014). ATO acknowledges support from the National Science Foundation of China (No. W2533010). LVD is supported by the Sherman Fairchild Postdoctoral Fellowship at the California Institute of Technology.

\bibliography{main}
\onecolumngrid
\appendix*
\setcounter{equation}{0}
\section*{End Matter}
\begin{figure*}[htbp!]
    \centering
    \includegraphics[width=\linewidth]{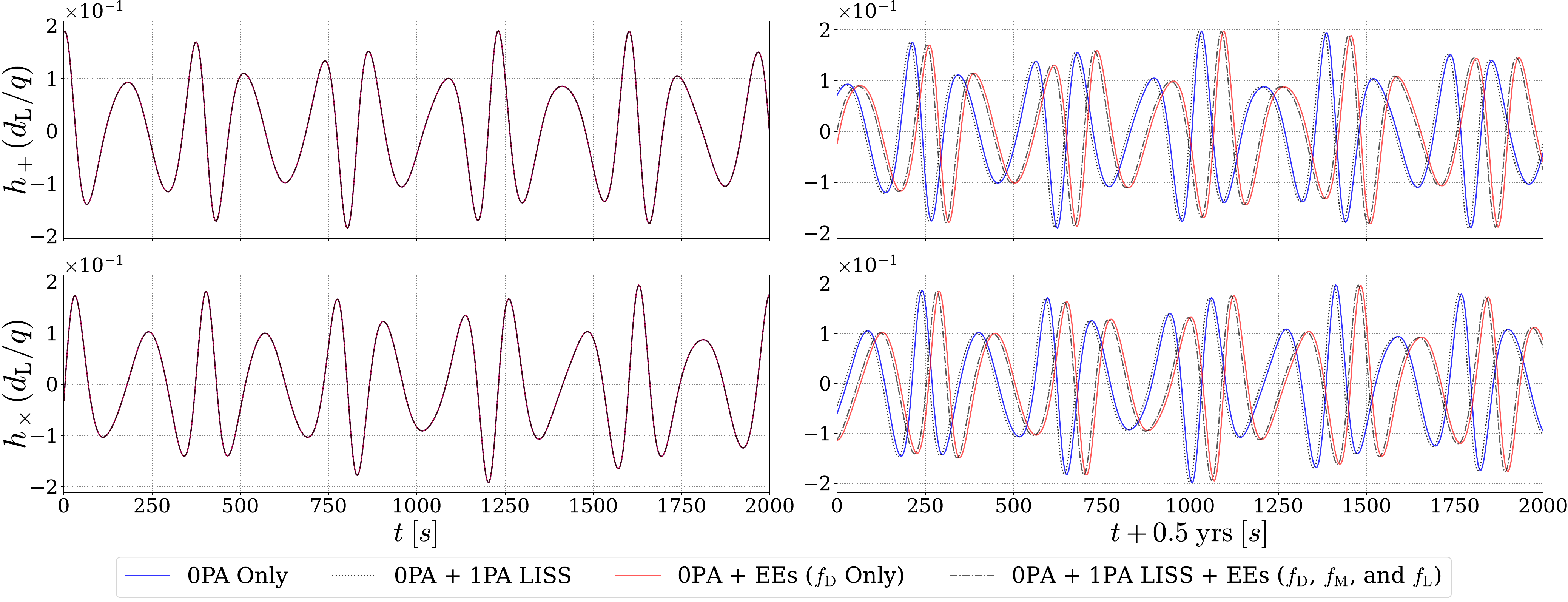}
    \caption{GW strain of an EMRI system ($q = 3 \times 10^{-5}$, $a = 0.9M$, $p_0 = 18M$, $e_0 = 0.25$) illustrating the accumulated dephasing due to spin-coupling effects and EEs in a co-rotating accretion disk. The top and bottom rows display the $+$- and $\times$-polarizations of the GW strain, respectively. The left column captures a $2000\,\mathrm{s}$ window at the beginning of the observation ($t=0$), and the right column captures an identical time window after $0.5$ years of orbital evolution. We compare four physical scenarios: the 0PA-order baseline (solid blue), 0PA evolution including LISS corrections (dotted black, $\chi=0.5$, $\beta=\pi/4$), EEs subject to aerodynamic drag but neglecting secondary spin (solid red) with gas density $\varepsilon = 10^{-12}$ ($\rho_{\mathrm{g}}\approx6.25\times 10^{-7}\,\mathrm{g \cdot cm^{-3}}$), and the fully inclusive evolution featuring both 1PA LISS and all spin-induced aerodynamic forces (dash-dotted black).}
    \label{fig:aero_GW_dephasing}
\end{figure*}
\twocolumngrid
\sect{Appendix A: Dephasing of Gravitational-Waves due to Environmental Effects}To model the EEs, we evaluate aerodynamic forces which arise from the total momentum transfer per unit mass ($\Delta\tilde{p}^i$) from the surrounding medium onto the secondary~\cite{Dyson_2024}
\begin{equation}
\begin{pmatrix} f_{\mathrm{D}} \\ f_{\mathrm{M}} \\ f_{\mathrm{L}} \end{pmatrix} = \gamma v \rho q^2 \begin{pmatrix} \Delta\tilde{p}^z \cos\beta - \Delta\tilde{p}^y \sin\beta \\ \Delta\tilde{p}^x \\ \Delta\tilde{p}^y \cos\beta + \Delta\tilde{p}^z \sin\beta \end{pmatrix},
\end{equation}
where $\gamma = (1-v^2)^{-1/2}$ and $v$ is the relative velocity of the secondary. In practice, rather than directly integrating the scattering geodesics to find the momentum transfer, these forces are rapidly evaluated using the polynomial interpolants provided in Eq. (53) of Ref.~\cite{Dyson_2024}. Their orbit-averaged fluxes are then incorporated into the orbital evolution as detailed in the main text (cf. Eq.~\eqref{averaged_fluxes}).

To obtain the complete GW signal, we first generate the EMRI inspiral trajectory by evolving the orbital constants of motion subject to the combined dissipative fluxes—specifically, the linear sum of the vacuum GW fluxes, the fluxes due to EEs, and the LISS fluxes. The baseline 0PA GW fluxes are computed strictly via the Teukolsky equation~\cite{Teukolsky1972, Teukolsky1973, Teukolsky1974}, and are efficiently evaluated by interpolating the \texttt{FEW} dataset~\cite{Chua_2019, Chua_2021, Chua_2021b, Chapman-Bird_2025}.
\begin{figure*}[htbp]
    \centering
    \includegraphics[width=\linewidth]{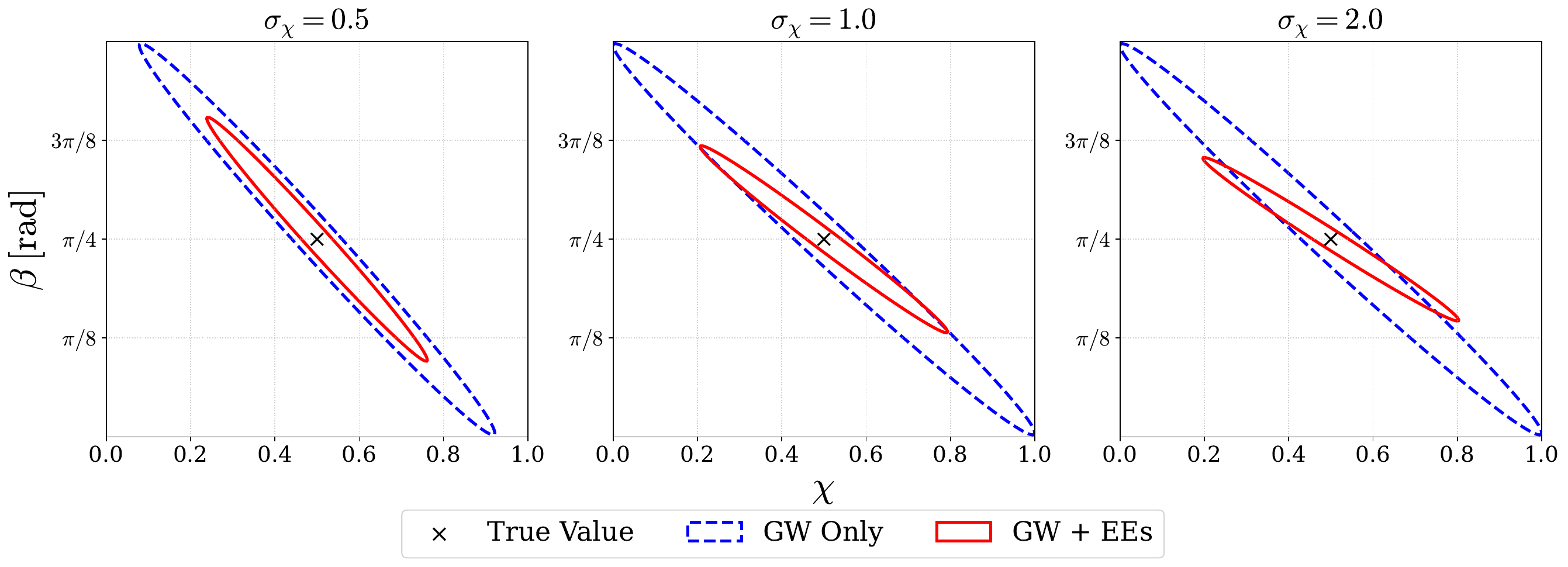}
    \caption{Projected $1\sigma$ marginalized Fisher constraints on the secondary spin magnitude $\chi$ and inclination angle $\beta$ for a 4-year EMRI observation ($q = 3 \times 10^{-5}$, $a = 0.9M$, $p_0 = 15M$, $e_0 = 0.25$). The three panels illustrate the sensitivity of the parameter estimation to the chosen prior standard deviation ($\sigma_{\chi} = 0.5, 1.0, 2.0$). Contours compare pure vacuum evolution (blue dashed) against an evolution incorporating EEs (red solid) within a co-rotating accretion disk, with density $\varepsilon =2.5 \times 10^{-13}$ ($\rho_\mathrm{g}\approx 1.56\times 10^{-7}\,\mathrm{g \cdot cm^{-3}}$). The true injected values ($\chi=0.5$, $\beta=\pi/4$) are marked by a black cross.}
    \label{fig:Fisher_comparison}
\end{figure*}

To accurately capture the dynamics from SCC of the extended secondary body, we supplement the trajectory with LISS GW fluxes. Following the formalism of Refs.~\cite{Skoupy_2023, Piovano_2026, Skoupy_2026}, the total secular rate of change for a constant of motion $\mathcal{C} \in \{\E, \AM, \K\}$ is expanded to linear order in the secondary's dimensionless parallel spin projection, $s_{\parallel}$, is $\dot{\mathcal{C}} = \dot{\mathcal{C}}^{(0)} + s_{\parallel}\dot{\mathcal{C}}^{(1)}$, where $\dot{\mathcal{C}}^{(0)}$ represents the standard geodesic contribution, while the 1PA spin correction $\dot{\mathcal{C}}^{(1)}$ is evaluated using the flux-balance laws on a linearly shifted virtual worldline gauge. By mapping the physical trajectory onto this virtual worldline, the 1PA source terms separate cleanly, allowing the spin-dependent fluxes to be expressed entirely in terms of asymptotic Teukolsky amplitudes and known geodesic functions. We compute these LISS flux corrections utilizing the  \texttt{KerrSpinningFluxes} package~\cite{Skoupy_2026}. Integrating these combined vacuum fluxes and EEs yields the secular time evolution of the quasi-Keplerian orbital parameters, $p(t)$ and $e(t)$. Throughout this analysis, we restrict the macroscopic orbit to the equatorial plane ($x_I=1$).  

The corresponding GW strain is constructed using a frequency-domain mode decomposition, where the two polarization states, $h_{+}$ and $h_{\times}$, are expressed as$$h_{+} - i h_{\times} = \sum_{\ell, m, n, k} \mathcal{A}_{\ell mnk} \, {^{-2}}S_{\ell m}(\vartheta, a\omega_{mnk}) e^{i(m\varphi - \omega_{mnk}t)},$$where $\mathcal{A}_{\ell mnk}$ denotes the complex Teukolsky amplitudes, ${^{-2}}S_{\ell m}$ are the spin-weighted spheroidal harmonics, and $\omega_{mnk} = m\omega_\varphi + n\omega_r + k\omega_\vartheta$ represent the fundamental mode frequencies~\cite{Drasco_2004, Drasco_2006}. Since both $\mathcal{A}_{\ell mnk}$ and $\omega_{mnk}$ are functions of the continuously evolving $p(t)$ and $e(t)$, evaluating the Teukolsky equation at every time-step during the multi-year trajectory is computationally prohibitive. To bypass this bottleneck, we adopt a rapid interpolation scheme analogous to the \texttt{FEW} framework~\cite{Chua_2019, Chua_2021, Chua_2021b}. We first precompute a dense $60\times 60$ grid of mode amplitudes and frequencies across the relevant $(p, e)$ parameter space using the \texttt{pybhpt} software package~\cite{Nasipak_2022, Nasipak_2024}. We then construct 2D interpolants using Chebyshev polynomials, verifying that the interpolation residual from an independent validation set is bounded by $\delta = 10^{-8}$ across the parameter space. We dynamically evaluate these interpolants along the integrated $[p(t), e(t)]$ tracks. To efficiently reconstruct the full EMRI signal with EEs, we sum the mode contributions following the truncation scheme in Sec. IVC of Ref.~\cite{Drasco_2006}, aggregating radial and polar harmonics until a strict fractional accuracy threshold of $\epsilon= 10^{-10}$ is achieved.

In FIG.~\ref{fig:aero_GW_dephasing}, we illustrate these combined physical effects on the GW strain. Though initially identical, the waveforms develop a visible dephasing over long timescales ($t \approx 0.5\;\mathrm{yrs}$) due to the accumulation of small secular flux changes. Late-time phases show that EEs are the primary driver of this dephasing, creating a pronounced shift between the vacuum (solid blue) and environmental (solid red) inspirals. Spin-coupling effects remain subdominant, causing only minor additional phase shifts when $\chi\neq0$ (dotted and dash-dotted black lines). 

\sect{Appendix B: Choice of Prior and its Effect on the Secondary's Spin Constraints}To demonstrate the data-driven nature of EMRIs with EEs, we perform parameter estimation across varying prior bounds on the secondary spin magnitude ($\sigma_{\chi} = 0.5, 1.0,$ and $2.0$) in FIG.~\ref{fig:Fisher_comparison}. The numerical results highlight a stark contrast in the likelihood surface between the GW- and EE-driven evolutions. In the pure vacuum scenario, the $1\sigma$ marginalized uncertainties scale dramatically with the injected prior. For instance, relaxing the prior from $\sigma_{\chi} = 0.5$ to $2.0$ causes the spin magnitude error to balloon from $\Delta \chi \approx \pm 0.422$ to $\pm 0.730$, while the inclination uncertainty similarly degrades from $\Delta \beta \approx \pm 0.843\,\mathrm{rad}$ to $\pm 1.459\,\mathrm{rad}$. Furthermore, the correlation between the two parameters remains pinned at $C_{\beta,\chi}=-1.000$ across all tested priors. This scaling confirms that the vacuum constraints are almost entirely prior-dominated; the data likelihood is flat along the degenerate projection $\chi \sin\beta$, causing the error ellipses to artificially expand until they saturate the boundaries imposed by the prior.

Conversely, the inclusion of EEs fundamentally alters this behavior, as the parameter constraints become more rigid and large changes in the prior produce small changes in the constraints for $\chi$ and $\beta$. Even when the prior standard deviation is quadrupled from $\sigma_{\chi} = 0.5$ to $2.0$, the $1\sigma$ uncertainty on the spin magnitude only widens marginally, shifting from $\Delta \chi \approx \pm 0.261$ to just $\pm 0.303$. The inclination error mirrors this stability, remaining tightly bounded between $\Delta \beta \approx \pm 0.526\;\mathrm{rad}$ and $\pm 0.610\;\mathrm{rad}$. While a strong correlation persists ($C_{\beta,\chi}\approx-0.998$), the geometrical degeneracy is broken. This stability confirms that in the regime where EEs are considerable, the measurement is strictly data-driven. The distinct functional dependence of the EEs imparts sufficient local curvature to the likelihood surface that the $\chi$ and $\beta$ are constrained directly by the signal itself, rather than by the artificial limits of the prior.
\end{document}